\documentclass[useAMS,usenatbib]{mn2e}
\usepackage{graphicx}
\usepackage{txfonts} 

\newcommand{\aap}{A\&A}
\newcommand{\aapr}{A\&A Rev.}
\newcommand{\mnras}{MNRAS}
\newcommand{\apj}{ApJ}
\newcommand{\aj}{AJ}

\newcommand{\apjs}{ApJS}
\newcommand{\araa}{ARA\&A}
\newcommand{\nat}{Nature}
\newcommand{\msait}{Mem. S.A.It.}
\newcommand{\pr}{Phys. Rep.}

\catcode`\@=11
\def\gsim{\ifmmode{\mathrel{\mathpalette\@versim>}}
    \else{$\mathrel{\mathpalette\@versim>}$}\fi}
\def\lsim{\ifmmode{\mathrel{\mathpalette\@versim<}}
    \else{$\mathrel{\mathpalette\@versim<}$}\fi}
\def\@versim#1#2{\lower 2.9truept \vbox{\baselineskip 0pt \lineskip
    0.5truept \ialign{$\m@th#1\hfil##\hfil$\crcr#2\crcr\sim\crcr}}}
\catcode`\@=12
\def\msun{\hbox{$M_\odot$}}

\def\yr-1{\hbox{${\rm yr}^{-1}$}}

\def\rsun{\hbox{$R_\odot$}}

\def\t9{\hbox{$t_9$}}

\def\m*{\hbox{$M_{\rm stars}$}}

\def\ho{\hbox{$H_\circ$}}
\def\h50{\hbox{$\ho /50$}}

\begin{document}

\title{A Transient Overcooling in the Early Universe? \\ Clues from Globular Clusters Formation}
\author[Alvio Renzini]{Alvio Renzini$^{1}$\thanks{E-mail: alvio.renzini@inaf.it}\\ 
$^{1}$INAF - Osservatorio
Astronomico di Padova, Vicolo dell'Osservatorio 5, I-35122 Padova,
Italy}

\date{Accepted 2023 June 8. Received 2023, June 5; in original form 2023, May 22 }
\pagerange{\pageref{firstpage}--\pageref{lastpage}} \pubyear{2002}

\maketitle
                                                            
\label{firstpage}

\begin{abstract}
The mere existence of multiple stellar generations in Milky Way globular clusters indicates that each generation was unable to stop star formation, that instead persisted unimpeded for several million years. This evidence argues  for an extended stage of star formation within a forming globular cluster, during which stellar feedback was substantially ineffective and the nascent globular cluster was able to accrete processed gas from its surrounding, and efficiently convert it into successive stellar generations. It has been argued that such delayed feedback  results from core collapse in most massive stars failing to trigger an energetic supernova explosion, but rather leading directly to black hole formation. Thus, globular clusters offer a concrete phenomenological example  for the  lack of feedback in young starbursts, an option that has been widely advocated  to account for the unexpected abundance of UV-luminous galaxies at $z=9-16$, as revealed by JWST observations.
The paper is meant to attract attention to this opportunity for a synergic cooperation of globular cluster and high redshift research.

\end{abstract}


\maketitle

\section{Introduction}
\label{sec:intro}
On its first Cycle, the {\it James Webb Space Telescope} (JWST) has revealed an unexpected abundance of UV-bright galaxies at $z=9-16$ (e.g., \citealt{finkelstein23, donnan23, mcleod23, harikane23a, harikane23b} and many others). Their finding was {\it unexpected} in the frame of existing theoretical models of galaxy formation and evolution, that had been fine tuned to reproduce observables in the lower redshift Universe, see e.g., Figure 14 in \cite{finkelstein23}.

As a consequence, a number of possible alternatives to current assumptions in cosmological simulations are being explored to cope with this discrepancy (e.g., \citealt{boyan23, wilkins23}). For example, \cite{yung23} discuss several options to ease the factor $\sim\!\!30$ underprediction of their semianalytical model 
at $z\sim\!\!13$ compared to the observed UV luminosity function. These include a top heavy IMF, that automatically would bust the UV luminosity function, exclude other possible limitations of their model such as a too inefficient baryon cooling within halos and inefficient gas-to-stars conversion. They  conclude that too {\it efficient stellar feedback} is the main  ingredient that could be responsible for their model underpredicting the number of luminous galaxies at these redshifts.  Indeed, the abundance of UV luminous galaxies appears to be consistent with the limiting case of (feedback unimpeded) $\sim\!\! 100$ per cent efficiency in converting baryons into stars in dark matter halos at these redshifts \citep{finkelstein23}, as estimated by \cite{behroozi18}.

In alternative, \cite{shen23} appeal to intrinsic UV luminosity variability of dwarf galaxies at high redshifts, such objects being likely dominated by a bursty mode of star formation. Thanks to a kind of intrinsic Eddington bias, the result of variability is to boost the top end of the luminosity function, with consistency with the observed counts being achieved for  a variability described by a gaussian luminosity spread with $\sigma_{\rm UV}\simeq 2.5$ mag.

A top heavy IMF is also favoured by \cite{harikane23a}, who, for lack of evidence, exclude AGN boosting the UV luminosity of $z\sim\!\!9-16$ galaxies. A lack of suppression of star formation by the UV background in the pre-ionization era is also mentioned as a possible contributor.

A specific case of feedback suppression at high redshifts is advocated by \cite{dekel23}. Based on the Starburst99 models \citep{leitherer99}, Dekel et al. argue that at low metallicities, such as those prevailing at high redshifts, stellar winds from massive stars convey little energy and momentum, hence prior to supernova (SN) explosions stellar feedback is very inefficient, hence boosting star formation and the UV luminosity function.
This low-feedback phase lasts up to $\sim\!\!3$ Myr since a burst of star formation, then coming to an end as the most massive stars eventually undergo SN explosions and efficient feedback begins. This assumes that all massive stars undergoing core collapse also end with a SN display, hence ejecting $\sim\!\!10^{51}$ erg in kinetic energy.

While all these options are being currently considered, this paper expands on the possibility of an even stronger reduction of stellar feedback, that can be achieved by substantially  extending the no-supernovae period, past a burst of star formation. The justification for such a scenario comes from the multiple stellar generation phenomenon in Galactic globular clusters (GC), which suggests a substantially more extended period of inefficient feedback, up to $\sim\!\!10$ Myr, as discussed  in \citet[hereafter Paper I]{renzini22}. Section \ref{sec:globulars} succinctly recaps the  evidence on GC formation supporting the concept of an extended feedback-free time at their formation. Section \ref{sec:overcooling} then expands on the consequences of such a delayed feedback for star formation in the early Universe, and finally, Section \ref{sec:final} returns on the key issues and on the plausibility of the whole scenario, including mentioning some caveats.

\section{Globular Cluster Formation}
\label{sec:globulars}
Virtually all (Galactic) GCs host multiple stellar generations, with the first generation (1G) reflecting the chemical composition of the ISM prior to the 1G formation, whereas second generation (2G) stars are depleted, to various degrees,  in carbon and oxygen and enriched in helium, nitrogen and sodium (e.g., \citealt{gratton12, milone17}).  Thus, the material having formed 2G stars had to be exposed to proton-capture processes at high temperatures in stars of the first generation, though their nature has been matter of debate ever since the multiple generation phenomenon  was discovered. In Paper I it is argued that the 
most likely candidates are massive interacting binaries, as originally suggested by  \cite{demink09}. The majority of massive stars are indeed members of interactive binaries \citep{sana12} and,  most of the nuclearly-processed material is shed with low kinetic energy  as a result of common-envelope events \citep{demink09}.

It is worth emphasising that the fraction of 2G stars increases from $\sim\!\!50$ per cent in lower mass GCs to over 80 per cent in the most massive ones \citep{milone17}. In all evidence, {\bf the formation of the first generation did not stop star formation, which actually had to continue with an even increasing rate and efficiency.} In Paper I, this absence of feedback was ascribed to a temporary lack of supernova explosions, as also required by most 2G stars having the same iron abundance of 1G ones, indicating no or very small contamination by supernova ejecta \citep[see also][]{milone17}. This also demonstrates that 2G stars had to form {\it before} SN explosions began to pollute the ISM, hence all star formation had to be confined within a few (up to $\sim\!\!10$) Myr. As discussed in Paper I, this can be due to stars more massive than $\sim\!\!20\,\msun$ failing to produce a SN display at their core collapse, but rather {\it silently} sinking into black holes \citep[see also][]{krause13}. Thus, following a burst of star formation there would be no supernova events for the first $\sim\!\! 10$ Myr, while stars more massive than $\sim\!\! 20\,\msun$ complete their evolutionary cycle and pollute the ISM with proton-capture products. Then supernovae begin and continue for another $\sim\!\! 25$ Myr, while stars from $\sim\!\! 20$ down to $\sim\!\! 8\,\msun$ complete their evolution. Strong feedback from these supernovae will then bring to an end star formation within the young GC. Finally,  no more core-collapse supernovae occur, from  $\sim\!\! 35$ Myr on  after the burst, when stars less massive than $\sim\!\! 8\,\msun$ end their evolution as white dwarfs.

In most GCs there appears to be a small 1G-2G discontinuity in chemical composition, suggesting a brief pause in star formation, then followed by a number of successive bursts, each generating a 2G sub-population. Such number increases with cluster mass \citep{milone17} and becomes as high as 15 in $\omega$ Cen \citep{bellini17}, the most massive Galactic GC. Clearly, even successive (2G) bursts were unable to stop star formation, before supernovae eventually succeeded. In all evidence,  feedback was ineffective also during most of the formation of the second generations.

During this no-supernova phase, may stellar winds still provide sufficient feedback to contrast  this scenario? At low metallicity winds carry little energy, as argued by \cite{dekel23}, but GCs exist with near-solar metallicity and they  still exhibit the multiple generation phenomenon \citep{kader22}. However, for the high ISM densities at GC formation, energy dissipation can be so high to neutralize most of the feedback from stellar winds, while radiative feedback or UV background may have only minor effects \citep{elmegreen17, dekel23}. In any event, the empirical fact remains, that, even in near-solar metalliciry GCs, the first generation did not prevent the formation of second generation stars.

The dominance of 2G stars demands that the 1G population having processed the material to form 2G stars had a substantially higher mass than that of the 1G still bound to the clusters. 
This is known as the {\it mass budget problem} (e.g., \citealt{renzini15}), whereby the first generation returns only $\sim\!\! 10$ per cent of its mass with the required composition to form 2G stars.
Hence, 1G stars still bound to the cluster today fall  short by at least a factor of $\sim\!\! 10$ in producing enough material to build the 2G stars. This mismatch would be even worse if star formation was restricted to the 3 Myr period prior to the first core-collapse event, as only the most massive stars would have had time to contribute, hence the necessity to extend longer the no-supernova phase, indicatively to $\sim\!\! 10$ Myr as justified in Paper I. As far as the mass of the 1G contributors is concerned, in Paper I it is argued that, within the (dwarf) galaxy hosting a  nascent GC (e.g., as seen at high redshifts, \citealt{vanzella19,vanzella23}), this extended 1G feeder population inhabited a wide  region  around the forming GC,  that was also actively star forming and feeding the central cluster with material to form the second generations.  During this {\it overcooling} phase, the young GC was actually the centre of a converging accretion flow, lasting some 10 Myr, before supernovae finally brought it to a halt. 

This extended feedback-free time is critical to ensure that a sufficient number of 1G donor stars, down to $\sim\!\!20\,\msun$, had time to evolve and shed enough processed materials to build the 2G stars in GCs.
Thus, the converging material had been processed inside  stars of a first generation that collectively was several times more
massive then the final mass of the bound GC (see also the GC formation models of \citealt{elmegreen17}).    It has been suggested that the young cluster R136, and its surrounding 30 Dor star-forming complex in LMC, may represent a local analog for GCs forming in high redshift  low-mass galaxies (\citet[Paper I]{schneider18}. 
It is worth emphasizing that the dominance of 2G stars requires that their formation had to take place with an extreme efficiency in gas to stars conversion, close to 100 per cent,  which was actually promoted by the lack of supernova feedback (Paper I). Clearly, if this is the way GCs formed, then the temporary lack of supernova explosions,  hence with the feedback delay promoting high star formation efficiency, ought to have important consequences for star formation in general, and possibly so for star formation in extremely high redshift galaxies.

In summary, the hypothesis of most massive stars ($\gsim 20\,\msun$) failing to result in a SN explosion has five decisive advantages for GC formation: 1) it promotes and extended, unimpeded formation of multiple stellar generations, 2) It boosts star formation efficiency, up to  $\sim\!\! 100$ per cent, with almost full conversion of gas into stars, 3)   it avoids contaminating 2G stars with heavy-element SN products, 4) it allows an extended range of stellar masses (i.e., above $\sim\!\! 20\,\msun$) to provide processed material for the formation of 2G stars, and 5) compared to the assumption of no delay in SN feedback, the extended period of star formation (from $\sim\!\! 3$ to $\sim \!\! 10$ Myr) allows a $\sim 30$ times bigger volume around the nascent GC to contribute material for the 2G formation (Paper I). As such, it appears to be promising to explore the consequences of this hypothesis for star formation in the early Universe.

\section{Overcooling in the Early Universe?}
\label{sec:overcooling}
\cite{white78} early noticed that cooling in the early Universe could be so effective that most baryons in dark matter halos would quickly turn into stars, an effect dubbed {\it overcooling}.
Clearly, global overcooling did not happen, because even by the present time not much more than 10 per cent of the cosmic baryons now reside in stars \citep{fukugita98}. Yet, if GCs formed as sketched above, with an early overcooling phase promoting the accumulation of multiple stellar generations in a short time interval, then an early overcooling as a generic properties of star formation may help accounting for the excess of UV-bright galaxies at $z\gsim 9$. Indeed, it offers an effective reduction of stellar feedback, invoked as the single simplest way of boosting star formation at very high redshifts, as from JWST observations \citep{finkelstein23, yung23, harikane23b, dekel23}. Indeed, compared to \cite{dekel23}, the proposed scenario extends from $\sim\!\!3$  to $\sim\!\! 10$ Myr  the ``feedback free" time past a burst of star formation. 

Besides this indirect role on the high-redshift Universe, where GC provide a hint favoring  delayed feedback and overcooling, young GCs and their immediate environment may also play a direct  role on global star formation at high redshift. High-redshift globulars and their precursors, their possible contribution to re-ionization and  observational detectability, have been considered for some time (see e.g., \citealt{carlberg02, schraerer11, katz13, trenti15, renzini17, boylan18, pozzetti19}). Thus, nascent GCs, with their 1G feeding surrounding, may contribute significantly to the stellar mass and luminosity  in the very early Universe.
Given the old ages of most Galactic GCs, they had to form beyond $z\sim\!\!3$, and the metal poor ones possibly well beyond it. Given the mass of GCs in the local Universe, young GCs, along with their $\sim\!\! 10$  times more massive star-forming environment, may have dominated over the whole stellar mass if formed beyond $z\sim\!\!5$ \citep{renzini17}. Also, given the extreme densities of typical GCs today, corresponding to some $\sim\! 10^7$ atoms cm$^{-3}$,  the  similarly high gas densities of forming GCs at very high redshift may have contributed significantly to the total emission measure in emission lines and nebular continuum. 

The object GN-z11 at $z=10.6$ \citep{bunker23} with density higher than $10^6$ cm$^{-3}$ has been proposed as (hosting) a possible GC in formation \citep{senchyna23, belokurov23}, though the high density may rather refer to the broad line region of an AGN \citep{maiolino23}. Worth noting is that its high N/O ratio is just what one expects for the  ISM of a globular cluster while forming its 2G stars, which indeed are strongly enhanced in nitrogen and depleted in oxygen.

\section{Discussion and Conclusions}
\label{sec:final}

It has been argued that GCs in the local Universe, with their ubiquitous multiple stellar generations,  offer strong evidence for a lack of feedback during their formation in the early Universe.
Stars  more massive  than $\sim\!\!20\,\msun$ failing to produce an energetic supernova would accomplish such effect, and delay feedback by some 10 Myr. If so, such delayed feedback and corresponding transient overcooling phase, would result in a major reduction in the stellar feedback, as now widely advocated as an obvious way of accounting for the excess of UV-bright galaxies at $z=9-16$ \citep{finkelstein23, dekel23, yung23}.

However, a few caveats are in order. For example, could such delayed feedback result in an overproduction of  stars  in cosmological simulations, not only in the early Universe but also all along cosmic times? It would certainly do so, but this could be compensated by suitably increasing the feedback efficiency past the $\sim\!\! 10$ Myr delay. In any event,  the overall star formation history may change little, but star formation would be more {\it bursty} on short timescales, and simulated galaxies may become more clumpy. It would be instructive to see how simulated galaxies would react to such a different recipe for feedback.

As already warned in Paper I, restricting core-collapse supernovae to below $\sim\!\!20\,\msun$ would have the collateral effect of reducing theoretical metal yields by roughly a factor of 2. Empirical yields, of course, would remain the same. Still, how physically-based is the assumption that the most massive stars fail to produce energetic SN events? This is actually a widely entertained 
possibility (e.g., \citealt{krause13,sukhbold16,adams17,sander19,eldridge22} and references therein). From the theoretical point of view, the opposite problem has, along the years, affected attempts of modelling  the outcomes of core-collapse events, namely the difficulty of producing explosions, especially in more massive stars \citep{sukhbold16}.

Ultimately, how well established is the empirical scenario of GC formation proposed in Paper I?  Like most previous attempts of describing how GCs may have formed, this is a phenomenological scenario, not a theoretical model. Yet, it is based on the enormous progress in the study of multiple stellar generations in GCs achieved mainly thank to HST imaging  and VLT spectroscopy. As such, it is built to comply, as well as possible, with all the resulting observational constraints.  In the end, for what matters for the feedback in starbursts, the mere existence of multiple stellar generations in GCs is tell tale evidence of inefficient feedback. Other details of the formation process may be irrelevant as far as the consequences for star formation at high redshifts are concerned.

Still, the empirical evidence of high, close to 100 percent efficiency in gas to star conversion refers to the scale of forming globular clusters, whereas the evidence for the $z=9-16$ Universe requires it at the scale of full young galaxies. A supernova avoidance period extended  to $\sim\!\! 10$ Myr would boost the star formation efficiency on all scales, but it remains to be seen whether this can reach close 100 per cent also on galactic scales at those redshifts. Nevertheless, supernova avoidance and ensuing overcooling favour the formation of high density  clumps, with further enhanced efficiency. The mentioned case of GN-z11, the best known galaxy at these redshifts, exhibits a very high gas density to the point of having the full  galaxy line emission being dominated  by  what has been proposed to be a globular cluster in formation.

\section*{Acknowledgments} 
I wish to thank Mauro Giavalisco for constructive comments and for his encouragement to write this paper.
I wish also to thank the anonymous reviewer for their questions that helped to improve the manuscript..

\section*{Data Availability}
No new data were generated or analysed in support of this research.

\author[0000-0002-7093-7355]{A.\,Renzini}

\label{lastpage}

\end{document}